\documentclass{ejm}

\usepackage{hyperref}

\usepackage[font=small]{caption}
\usepackage{amsmath}
\usepackage{amssymb}
\usepackage{graphicx}
\usepackage[letterpaper, margin=1in]{geometry}
\usepackage{subfig}
\usepackage{mathrsfs}
\usepackage{gensymb}

\def\barr{\begin{array}}
\def\earr{\end{array}}

\def\vu{\vec{u}}
\def\beq{\begin{equation}}
\def\eeq{\end{equation}}

\let\realverbatim=\verbatim
\let\realendverbatim=\endverbatim
\renewcommand\verbatim{\par\addvspace{6pt plus 2pt minus 1pt}\realverbatim}
\renewcommand\endverbatim{\realendverbatim\addvspace{6pt plus 2pt minus 1pt}}

\ifprodtf \else
  \checkfont{eurm10}
  \iffontfound
    \IfFileExists{upmath.sty}
      {\typeout{^^JFound AMS Euler Roman fonts on the system,
                   using the 'upmath' package.^^J}%
       \usepackage{upmath}}
      {\typeout{^^JFound AMS Euler Roman fonts on the system, but you
                   don't seem to have the}%
       \typeout{'upmath' package installed. EJM.cls can take advantage
                 of these fonts,^^Jif you use 'upmath' package.^^J}%
       \providecommand\mathup[1]{##1}%
       \providecommand\mathbup[1]{##1}%
      }
  \else
    \providecommand\mathup[1]{#1}%
    \providecommand\mathbup[1]{#1}%
  \fi
\fi

\ifprodtf \else
  \checkfont{msam10}
  \iffontfound
    \IfFileExists{amssymb.sty}
      {\typeout{^^JFound AMS Symbol fonts on the system, using the
                'amssymb' package.^^J}%
       \usepackage{amssymb}%
         \let\leq=\leqslant
         \let\geq=\geqslant
      }{}
  \fi
\fi

\ifprodtf \else
  \IfFileExists{amsbsy.sty}
    {\typeout{^^JFound the 'amsbsy' package on the system, using it.^^J}%
     \usepackage{amsbsy}}
    {\providecommand\boldsymbol[1]{\mbox{\boldmath $##1$}}}
\fi

\newsavebox{\astrutbox}
\sbox{\astrutbox}{\rule[-5pt]{0pt}{20pt}}

\newdefinition{definition}[theorem]{Definition}

\usepackage{color}

\title[European Journal of Applied Mathematics]{Deformation of an Elastic Substrate Due to a Resting Sessile Droplet}

\author[Bardall et al.]{%
Aaron Bardall$^1$, 
Karen E. Daniels$^2$,
\and Michael Shearer$^1$
}

\affiliation{%
  $^1\,$Dept. of Mathematics, N.C. State University, Raleigh, NC 27695 US\\
    email\textup{\nocorr: \texttt{arbardal@ncsu.edu}}\\
  $^2\,$Dept. of Physics, N.C. State University, Raleigh, NC 27695 US}

\date{29 June 2016}
\pubyear{2016}
\volume{000}
\pagerange{\pageref{firstpage}--\pageref{lastpage}}

\begin{document}

\label{firstpage}
\maketitle

\begin{abstract}
On a sufficiently-soft substrate, a resting fluid droplet will cause significant deformation of the substrate. 
This deformation is driven by a combination of capillary forces at the contact line and the fluid pressure at the solid surface.  These forces are balanced at the surface by the solid traction stress induced by the substrate deformation. 
Young's Law, which predicts the equilibrium contact angle of the droplet, also indicates an {\textit{a priori}} radial force balance for rigid substrates, but not necessarily for soft substrates which deform under loading.  It remains an open question whether the contact line transmits a non-zero radial force to the substrate surface in addition to the conventional vertical force.
We present an analytic Fourier transform solution technique that includes general interfacial energy conditions which govern the contact angle of a 2D droplet. This includes evaluating the effect of gravity on the droplet shape in order to determine the correct fluid pressure at the substrate surface for larger droplets. Importantly, we find that in order to avoid a strain singularity at the contact line under a nonzero radial contact line force, it is necessary to include a previously-neglected radial traction boundary condition.  To quantify the effects of the contact line and identify key quantities that will be experimentally-accessible for testing the model, we evaluate solutions for the substrate surface displacement field as a function of Poisson's ratio and zero/non-zero radial contact line forces. 
\end{abstract}


\section{Introduction}

The motion of droplets across substrates is crucial to droplet-based microfluidics and micro-fabrication \cite{lin07}. The means for controlling such motions is highly varied, and includes temperature gradients \cite{onuki05}, magnetic fields \cite{gao04}, and surface chemistry \cite{chaudhury92}.  Though these effects are well-quantified for rigid substrates, softer materials such as hydrogels and biological tissues are themselves deformed  by contact with fluid droplets \cite{andreotti16,style16}.  Consequently, these softer materials can experience significant strain due to capillary forces acting at the droplet contact line so that the onset of droplet motion and the subsequent dynamics may be quite different from the corresponding behavior on a rigid substrate.

Before addressing droplet dynamics on soft substrates, it is necessary to first consider the simpler case of a symmetrical droplet at rest on a  soft substrate.  Analytical expressions for displacement fields within neutrally wetted substrates (contact angle of $90^{\circ}$) have previously been determined \cite{jerison11,limat12,style12}.  However, since droplet motion can result from non-uniform contact angles, there is a need for a model that allows partial wetting (contact angle $\neq 90^{\circ}$); such a model was developed and analyzed recently \cite{bostwick14}. Here, we present a similar analysis for a two dimensional droplet.  This simplification has the benefit that it allows for the consideration of different contact angles at the front and back contact line, without the complication of a varying contact angle around the perimeter of a three-dimensional droplet.  In addition, this two-dimensional analysis reveals  that in order to balance forces with finite strains, an additional radial traction boundary condition must be included. We show that this new boundary condition is sufficient to regularize the radial strain, which would otherwise have a singularity at the contact line. In what follows, we use the term {\em radial} to refer to the horizontal variable in two dimensions. We also use the three-dimensional terminology {\em contact line,} although in two dimensions this is simply a pair of triple points at which the substrate and droplet surfaces are both in contact with the atmosphere.

For the case of a hard substrate, the radial force balance at the contact line is governed by the Young-Dupr\'e equation with equilibrium droplet contact angle $\alpha$:
\beq\label{young1}
\gamma_{sg} - \gamma_{ls} = \gamma \cos\alpha,
\eeq
where $\gamma$ represents the surface energy for each phase interface ($sg$ for solid-gas and $ls$ for liquid-solid; $\gamma=\gamma_{lg}$ represents the liquid-gas surface energy of the droplet).   For soft elastic substrates, the surface energy $\gamma$ may differ significantly from the surface stress $\Upsilon$.  A proposed relation between these terms is the Shuttleworth equation \cite{marichev11} $\Upsilon=\gamma+\partial\gamma/\partial\varepsilon,$ in which $\varepsilon$ is the strain parallel to the interface.  
This difference between surface energy $\gamma$ and surface stress $\Upsilon$ allows for 
the generation of radial contact line forces. To account for this possibility, we use a generalized contact line force law 
\cite{weijs14} for droplets whose contact angle is at equilibrium. In addition, the vertical component of the contact line force (which plays no role on a hard substrate) induces a vertical deformation on the order of the elastocapillary length scale $L_e = \gamma/E$ at the contact line, where $E$ represents the elastic modulus of the substrate. Consequently, there is significant substrate deformation in both radial and vertical directions near the contact line.

Previous work \cite{bostwick14,jerison11,style12} included the solid surface stress to regularize a vertical strain singularity at the contact line.  It was shown \cite{jerison11} that including the surface traction caused by the free surface shape is necessary for the transformed vertical displacement to decay sufficiently in Fourier space to provide a bounded displacement at the contact line location.  This surface traction has previously been estimated \cite{bostwick14,jerison11,style12} by linearizing the curvature of the substrate surface in the vertical direction and neglecting the radial component of curvature.  Though this is sufficient to regularize and provide accurate results for the vertical strain, we show that it does not regularize the radial strain under the generalized contact line force.  In this paper, we provide an estimate for the radial component of the surface traction and show that the strain is regularized under our new radial traction boundary condition, avoiding what otherwise would be an unphysical singularity in the radial strain at the contact line.  Moreover, inclusion of the radial traction boundary condition likely increases the accuracy of the radial deformation calculations with or without a radial contact line force, though experimental results for radial deformation are not currently available for comparison.

In addition to considering partial wetting and a general radial contact line force, our model can incorporate the effect of gravity on the droplet shape.  In the absence of gravity, the droplet minimizes its surface energy by assuming a circular shape. The presence of gravity flattens this circular shape, but this effect is negligible for droplets with radii much smaller than the capillary length $L_c = \sqrt{\gamma/\rho g}$, where $\gamma$ and $\rho g$ are the surface tension and specific weight of the droplet respectively.  The droplet curvature determines the Laplace pressure in the droplet, while the height of the droplet determines the hydrostatic pressure at the surface of the substrate.  These two pressures combined then influence the substrate displacement.  Previous work \cite{gomba12} included an analytical solution for a  droplet with shallow contact angle on a solid substrate.  Our solution includes contact angles up to $90^\circ,$ where the droplet shape can be expressed as a function in cartesian coordinates, neglecting disjoining pressure.  The fluid pressure is then determined, and contributes to surface forces on the substrate.

The paper is organized as follows.  In \S\ref{sec:problem} we calculate the variational derivation of droplet shapes in the presence of gravity, and the associated pressure   at the substrate surface. We then develop equations describing the static deformation within the substrate, and the associated boundary conditions. In \S\ref{sec:potfunc} we introduce a potential function $\psi$, which has the effect of reducing the analysis to that of a scalar equation. The deformation and stress are then expressed as linear combinations of derivatives of $\psi$.  The scalar equation is analyzed in Fourier space in \S\ref{sec:fourier}, leading to expressions for the surface deformations as integral equations.  The solution to the equations in \S\ref{sec:fourier} is then approximated in \S\ref{sec:surfdef} by use of an asymptotic expansion of the transformed functions and truncation of the inverse transform integrals, with the procedure being justified with the error analysis of \S\ref{sec:error}. (Details of this approximation are included in Appendix~\ref{sec:curvature} and supplementary materials.) Numerical results are presented in \S\ref{sec:results}, and a discussion in \S\ref{sec:discuss} concludes the paper.

\section{Problem Setup \label{sec:problem}}

\begin{figure}
\centerline{
\includegraphics[scale=0.8]{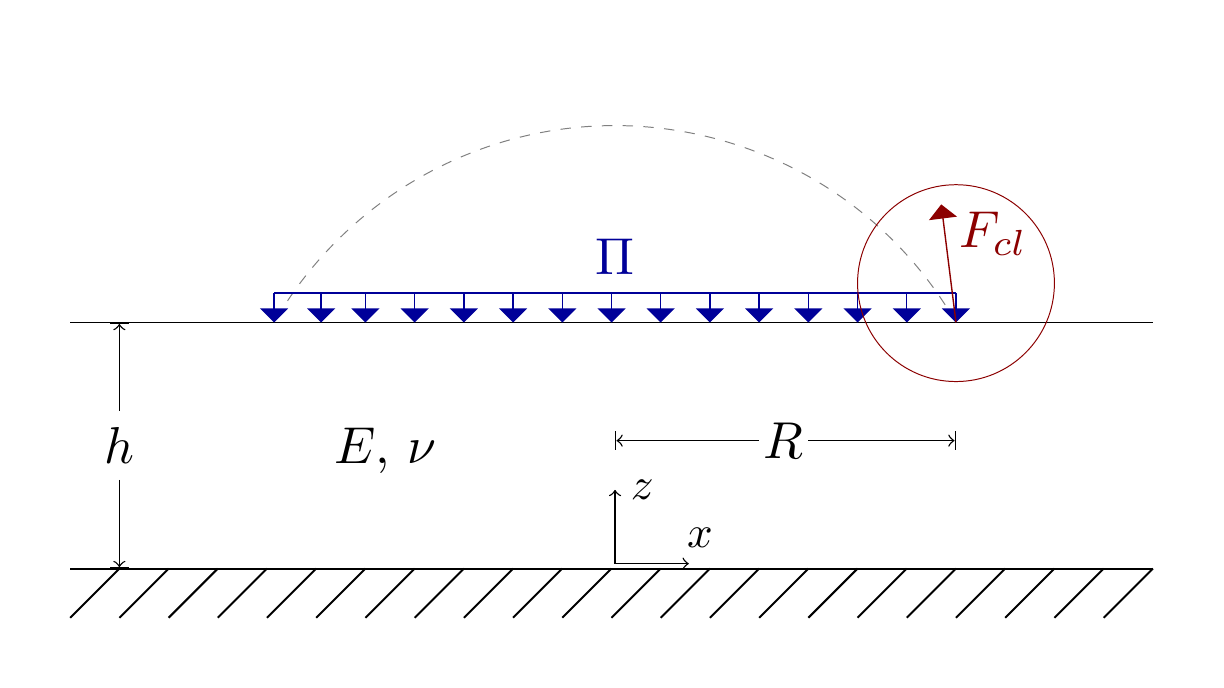}
\includegraphics[scale=0.8]{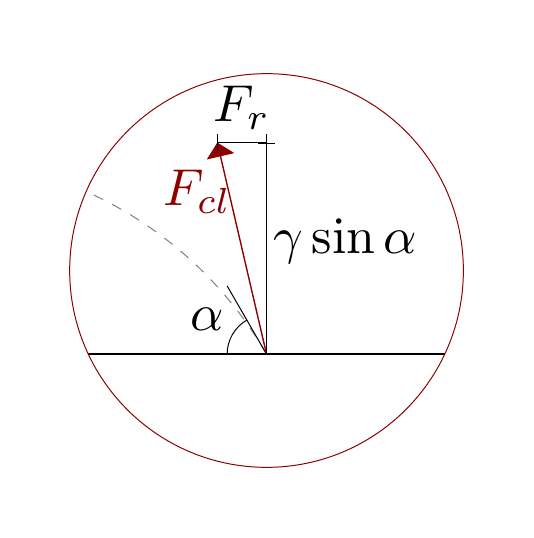}}
\caption{Pressure $\Pi$ and contact line force $F_{cl}$ acting on a stiff substrate with elastic modulus $E$ and Poisson's ratio $\nu$. The dashed curve represents the surface of the resting sessile droplet. In the blow-up near the contact line on the right, the contact line force includes the possibility of a nonzero radial stress component $F_r,$ discussed in \S\ref{sec:model}.}
\label{fig1}
\end{figure}

We consider a two dimensional droplet, depicted schematically in Fig.~\ref{fig1}, with width $2R$ and resting on the free upper surface of a soft elastic substrate.  In the reference configuration (no droplet), the substrate is taken to be fixed on the bottom surface $z=0,$ to have infinite extent, and to have constant thickness $h$. The elastic modulus of the substrate is denoted by $E$,  and  Poisson's ratio by $\nu$.   The contact line creates a vertical force ($\gamma\sin\alpha$) and a radial force ($F_r$) which cause significant deformation in a neighborhood of the contact line.   The fluid pressure $\Pi$ in the droplet acts at the substrate interface to compress the substrate below.  In this paper, we quantify these influences and describe the deformation of the substrate.  The deformation is analyzed for both the conventional contact line model which assumes no net radial contact line force ($F_r = 0$), and the generalized contact line model quantified by Weijs \textit{et al} \cite{weijs14} (discussed in \S\ref{sec:model}) which has a non-zero $F_r$.

The model depends largely on the formulation of boundary conditions at the free surface of the substrate. This is composed of two parts; the section under the droplet, and the solid-gas interface between the substrate and air.   The effect  of the droplet is expressed solely through the surface stress at the substrate surface and through the pressure $\Pi$. Once these are quantified, the droplet is effectively removed from the subsequent analysis.

To determine the shape of the substrate free surface, we formulate a boundary value problem for the elastic displacement within the substrate. It is convenient to use Eulerian coordinates $(x,z),$ shown in Fig.~\ref{fig1}. 
in which the substrate free boundary is  located at $z=h$  in the reference configuration. The displacement $\vu$ of the substrate   is then represented in two components by
\beq
\vu(x,z) = u(x,z) \hat{e}_x + w(x,z) \hat{e}_z, \ \ -\infty<x<\infty, \ 0<z<h,
\eeq
where $\hat{e}_x$, $\hat{e}_z$ are unit vectors in the coordinate directions. The displacement is defined relative to the reference configuration, mapping the reference configuration to the static deformed substrate configuration:
\[ \big(x,z\big) \mapsto \big(x+u(x,z),z+w(x,z)\big). \]

\subsection{Droplet Shape and Fluid Pressure}

The surface pressure $\Pi$ and droplet shape are influenced only slightly by the deformation in the substrate, which is localized near the contact line. In this section, we determine the pressure and droplet shape by assuming the substrate is rigid and flat. With this assumption, we determine the relationship between the parameters $\Pi$ and $R,$ and their dependence on surface tension and gravity.

Gravity influences droplets when the droplet size exceeds the capillary length scale: $R> L_c$.  For low capillary numbers ($R/L_c \ll 1$), the droplet surface  takes on a circular   shape (spherical in three dimensions).  For large capillary numbers, gravity dominates and the droplet flattens out except near the contact line.  

The height $f(x)$ of the droplet free surface above the substrate is determined by minimizing the total energy.  The differential gravitational and surface potential energies are  given respectively by
\beq\label{diff_energy}
dU_g(x) = \frac{\rho g f(x)^2}{2} dx \qquad dU_s(x) = \gamma\sqrt{1+f'(x)^2} dx.
\eeq
We then consider the energy cost functional $U$ representing the energy of half the droplet ($0\leq x\leq R$), imposing a constant area $A$ representing the amount of fluid in the droplet:
\beq\label{energy_funct}
U(f)= \int dU_g + dU_s - \lambda dA = \int_0^R \big[\rho gf(x)^2/2 + \gamma\sqrt{1+f'(x)^2} - \lambda f(x)\big] dx
\eeq
where $\lambda$ is a Lagrange multiplier.  
The corresponding Euler-Lagrange equation 
results in the ODE
\beq\label{shape_ode}
\lambda =\rho gf(x) -\gamma\frac{f''(x)}{(1+f'(x)^2)^{3/2}} = \Pi_\text{hydrostatic} + \Pi_\text{Laplace} = \Pi.
\eeq

From this we conclude that the pressure distribution under the droplet is constant,   the value of the Lagrange multiplier $\lambda$.  A low gravity pressure approximation is then obtained assuming a circular droplet profile:
\beq\label{lowgrav}
\Pi = \frac{\gamma\sin\alpha + \rho gA}{R} \sim \frac{\gamma\sin\alpha}{R} + 
\frac{\rho gR (\alpha\csc^2\alpha - \cot\alpha)}{2}, \qquad \text{as } R/L_c \to 0.
\eeq

The differential equation \eqref{shape_ode} is   solved more generally by exploiting the chain rule and imposing boundary conditions $f'(0) = f(R) = 0$, $f'(R) = -\tan \alpha$ to provide an implicit solution for the droplet shape in terms of the non-dimensionalized pressure $\Pi/\rho gL_c$ and contact angle $\alpha:$
\beq\label{shape_sol}
\frac{R-x}{L_c} = \int_0^{f(x)/L_c} \frac{ d\xi}{\sqrt{\big(\frac{1}{2}\xi^2 - \frac{\Pi}{\rho gL_c}\xi - \cos\alpha\big)^{-2} - 1}}, 
\eeq
where the peak droplet height, determined using the boundary condition $f'(0)=0,$ is given by
\beq\label{shape_ic}
\frac{f(0)}{L_c} = \frac{\Pi}{\rho gL_c} - \sqrt{\Big(\frac{\Pi}{\rho gL_c}\Big)^2 - 2\big(1 - \cos\alpha\big)}.
\eeq

\begin{figure}
\centerline{\includegraphics[scale=0.5]{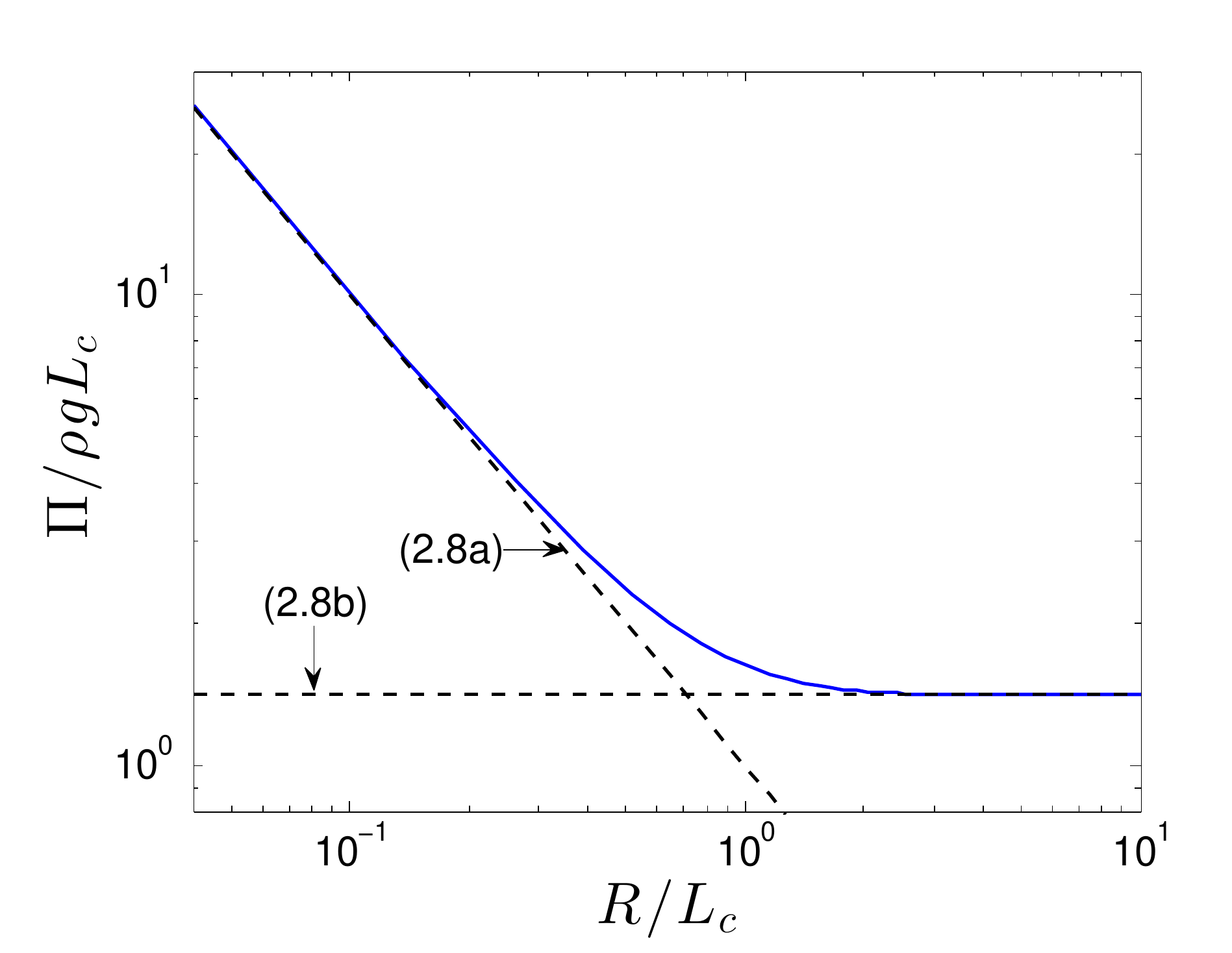}}
\caption{Nondimensionalized fluid pressure $\Pi/\rho gL_c$ as a function of droplet radius for contact angle $\alpha = 90^\circ$.}
\label{loglog}
\end{figure}

For each choice of the dimensionless parameter $\Pi/\rho gL_c,$ we see that letting $x=0$ in \eqref{shape_sol} we obtain a value of $R/L_c,$ with the upper limit in the integral given by \eqref{shape_ic}. In Fig.~\ref{loglog}, we plot the curve of such values for a contact angle of $\alpha=90^\circ$ obtained numerically from \eqref{shape_sol}, \eqref{shape_ic} for specified $R/L_c$.  Once the pressure is calculated, the droplet shape can be obtained implicitly from \eqref{shape_sol}.

As shown in Fig.~\ref{loglog}, there is a clear transition from the capillary regime $R/L_c \ll 1$ to the gravitational regime $R/L_c \gg 1$.  The limiting regimes are determined analytically to be
\beq\label{pressure_lim}
\begin{array}{cclr}
\Pi \sim \gamma\sin\alpha/R & & \text{Capillary Regime} \qquad\qquad &(a)\\
\Pi \sim 2\rho gL_c\sin\frac \alpha 2 & & \text{Gravitational Regime}&(b)
\end{array}
\eeq
A justification of the gravitational pressure limit (\ref{pressure_lim}b) is provided in Appendix~\ref{sec:gravity}.

\subsection{Model equations and boundary conditions \label{sec:model}}

 The elastostatic Navier equations
\begin{equation}\label{navier}
(1-2\nu)\Delta\vu + \nabla(\nabla \cdot \vu) = 0,
\end{equation}
express force balance within the substrate.
Here $\nu$ is the Poisson ratio of the substrate, where incompressible solids have a Poisson ratio of $\nu = 1/2$.  In two dimensions, the strain $\varepsilon$ and stress $\tau$ are represented by $2\times 2 $ matrices with components,
\begin{equation}
\varepsilon_{ij} = \frac{1}{2} \Big[\frac{\partial{u_i}}{\partial{x_j}} + \frac{\partial{u_j}}{\partial{x_i}}\Big]
\end{equation}
and
\begin{equation}
\tau_{ij} = \frac{E}{1+\nu}\Big[\varepsilon_{ij} + \frac{\nu}{1-2\nu} \delta_{ij}\big(\varepsilon_{11}+\varepsilon_{22}\big)\Big]
\end{equation}
where $(u_1,u_2)$ correspond to displacements $(u,w)$ and spatial variables $(x_1,x_2) = (x,z)$.  In these tensors, $E$ represents the elastic modulus of the substrate and $\delta_{ij}$ is the Kronecker delta.

Boundary conditions are set at the solid surface $z=0,$ where the substrate has no displacement:
\begin{equation}\label{bcs1}
(u,w)|_{z=0} = (0,0). 
\end{equation}
The effect of the droplet on the substrate is quantified by defining the shear stress $\tau_{xz}$ and normal stress $\tau_{zz}$ at the free surface $z=h$:
\begin{equation}\label{bcs23}
(\tau_{xz},\tau_{zz})|_{z=h} = F_x(x)\hat{e}_x+F_z(x)\hat{e}_z + \Upsilon(x)\vec{\kappa}(x).
\end{equation}
Here $F_x(x)$ and $F_z(x)$ include the general contact line force $F_{cl}$ located at $|x|=R$ with radial component $F_r$ specified below and vertical component $\gamma\sin\alpha$ shown in Fig.~\ref{fig1}, as well as the fluid pressure $\Pi$.  The final term in \eqref{bcs23}, $\Upsilon(x)\vec{\kappa}(x)$, defines the traction stress generated by the resulting shape of the deformed free surface, where $\vec{\kappa}(x)$ is the curvature vector of this surface.  By considering the general curvature vector instead of just the linearized vertical component, as done in previous work \cite{bostwick14,jerison11,style12}, we will show that the strain is bounded at the contact line in the radial as well as the vertical directions under a general contact line force.  The solid surface stress is represented as the piecewise constant function 
\beq\label{ups1}
\Upsilon(x) = \Upsilon_{sg} + \Delta\Upsilon H(R-|x|), \ \ \mbox{with} \ \ 
\Delta\Upsilon = \Upsilon_{ls} - \Upsilon_{sg},
\eeq
in which   $H$ is the Heaviside function.   
The vector $\vec{r}(x) = \langle x+u,w+h\rangle |_{z=h}$ parameterizes the substrate free surface. Then the curvature vector is given as
\beq\label{curvature}
\vec{\kappa}(x) = \frac{(1+\partial_xu)\partial_{xx}w - \partial_{xx}u\partial_xw}{\big((1+\partial_xu)^2+(\partial_xw)^2\big)^2}\big((-\partial_xw) \hat{e}_x + (1+\partial_xu) \hat{e}_z\big)\big|_{z=h}.
\eeq

The conventional model assumes no radial contact line force ($F_x \equiv 0$).  In this case, which simplifies the model, there is a bounded solution at the contact line.  The inclusion of an approximation to the radial component of curvature in \eqref{bcs23} is expected to improve the fidelity of the solution.  In the general model ($F_x \neq 0$), the inclusion of the radial curvature approximation ensures a bounded solution, which otherwise would experience a singularity at the contact line.  Here, we allow the generality considered in previous work \cite{bostwick14} by defining the radial stress as
\begin{equation}\label{fx}
F_x(x) = -F_r\delta(R-|x|)\text{sgn}(x),
\end{equation}
where $\delta$ is the Dirac delta function.  In the generalized model, this radial contact line force $F_r$ arises due to the difference between surface stress and surface energy, and in general is represented as $F_r = (\Upsilon_{ls} - \Upsilon_{sg}) - (\gamma_{ls} - \gamma_{sg})$.  
The form proposed by Weijs {\it et al},
\beq\label{J}
F_{r} =\frac{1-2\nu}{1-\nu}\gamma(1+\cos\alpha),
\eeq
  is derived in detail \cite{weijs14}.  Combining the vertical force at the contact line with the fluid pressure $\Pi$ acting under the droplet, we have
\begin{equation}\label{fz}
F_z(x) = \gamma\sin(\alpha)\delta(R-|x|) - \Pi H(R-|x|).
\end{equation}

\subsection{Potential Function \label{sec:potfunc}}

It is useful to express the displacement vector $\vu(x,z)$  in terms of a potential function $\psi$.  To do this, we define a Galerkin vector $\mathbf{G} = \psi(x,z)\hat{e}_z,$ 
and let
\begin{equation}
\vu(x,z) =2(1-\nu)\Delta\mathbf{G} - \nabla(\nabla\cdot \mathbf{G}).
\end{equation}
Substituting this into \eqref{navier}, we find that $\psi$ is biharmonic:
\begin{equation}\label{biharmonic}
\Delta^2\psi(x,z) = 0.
\end{equation}
The spatial scaling
\[ x = R\tilde{x}, \qquad z= R\tilde{z} \]
is applied to non-dimensionalize the radial distance.  The tilde's are dropped from the spatial variables, and the free surface of the substrate is now located at $z = \tilde{h} = h/R$.  We now reformulate the displacements and stresses component-wise in terms of the potential function $\psi$:
\begin{subequations}\label{dt}
\begin{eqnarray}
u &=& -\frac{1}{R^2}\frac{\partial^2{\psi}}{\partial{x}\partial{z}}, \label{dt1}\\
w &=& \frac{2(1-\nu)}{R^2}\frac{\partial^2{\psi}}{\partial{x}^2} + \frac{(1-2\nu)}{R^2}\frac{\partial^2{\psi}}{\partial{z}^2},\label{dt2}\\
\tau_{xz} &=& \frac{E}{(1+\nu)R^3}\Big((1-\nu)\frac{\partial^3{\psi}}{\partial{x}^3} - \nu\frac{\partial^3{\psi}}{\partial{x}\partial{z}^2}\Big),\label{dt3}\\
\tau_{zz} &=& \frac{E}{(1+\nu)R^3}\Big((2-\nu)\frac{\partial^3{\psi}}{\partial{x}^2\partial{z}} + (1-\nu)\frac{\partial^3{\psi}}{\partial{z}^3}\Big).\label{dt4}
\end{eqnarray}
\end{subequations}

\section{Solving the equations \label{sec:solve}} 
 
In this section, we first manipulate the biharmonic equation \eqref{biharmonic} using a Fourier transform and obtain a system of equations to solve for the transformed potential.  The Fourier transform is conducive to the cartesian coordinate representation of our two dimensional droplet, whereas a Hankel transform is appropriate in axisymmetric coordinates for the three dimensional droplet \cite{bostwick14}.   The surface displacements are recovered by truncating the inverse transform at large wave numbers.  The asymptotics of the error in this approximation  show that the strain is completely regularized  when using both traction boundary conditions.

\subsection{Fourier Transform Representation \label{sec:fourier}}

We define the Fourier Transform pair for $\psi$ as
\begin{equation}
\mathcal{F}(\psi) = \hat{\psi}(s,z) = \frac{1}{\sqrt{2\pi}}\int_{-\infty}^\infty \psi(x,z)e^{isx}dx, \qquad \psi(x,z) = \mathcal{F}^{-1}(\hat{\psi}) = \frac{1}{\sqrt{2\pi}}\int_{-\infty}^\infty \hat{\psi}(s,z)e^{-isx}ds
\end{equation}
Then \eqref{biharmonic} transforms to
\begin{equation}\label{fteqn}
 \Big(\frac{d^2}{dz^2} - s^2\Big)^2\hat{\psi} = 0.
\end{equation}
Solving \eqref{fteqn} with wave number $s$ as a parameter yields
\begin{equation}\label{psi1}
\hat{\psi}(s,z) = \big(A(s) + szB(s)\big)\cosh(sz) + \big(C(s) + szD(s)\big) \sinh(sz).
\end{equation}
Transforming (\ref{dt}a,b), and then applying the boundary conditions \eqref{bcs1}, we find  
\beq\label{coeffs1}
\begin{array}{ccccc}
A(s) = 2(1-2\nu)D(s),\qquad B(s) = -C(s).
\end{array}
\eeq
These expressions are substituted into \eqref{psi1} where the transformed potential $\hat{\psi}$ is now an expression involving two unknown Fourier coefficients $C(s)$ and $D(s)$.  These are determined from the two stress boundary conditions \eqref{bcs23} as follows. Recall that the transformed stresses, from  (\ref{dt}c,d), are now linear  combinations of  $C(s)$ and $D(s)$. 

Transforming the shear stress boundary condition from \eqref{bcs23} we obtain
\[ \hat{\tau}_{xz}|_{z=\tilde{h}} = \mathcal{F}(F_x) + \mathcal{F}\big(\Upsilon(x)(\vec{\kappa}\cdot\hat{e}_x)\big)|_{z=\tilde{h}}. \]
We define the function $M(s)$  using  \eqref{fx}, by
\[ iM(s) = \mathcal{F}(F_x) = \frac{-2F_ri\sin s}{\sqrt{2\pi}R}. \] 
The convolution identity $\mathcal{F}\big(f(x)g(x)\big) = \mathcal{F}\big(f(x)\big)*\mathcal{F}\big(g(x)\big)/\sqrt{2\pi}$, and \eqref{ups1} lead to the first equation defining the unknowns $C(s), D(s):$
\begin{equation}\label{tauxz_1_alt}
-i\Big(\hat{\tau}_{xz} - \Upsilon_{sg}\mathcal{F}(\vec{\kappa}\cdot\hat{e}_x) - \frac{\Delta\Upsilon}{\pi}\frac{\sin s}{s}*\big(\mathcal{F}(\vec{\kappa}\cdot\hat{e}_x)\big)\Big)\Big|_{z=\tilde{h}} = M(s).
\end{equation}
Similarly, transforming the normal stress boundary condition from \eqref{bcs23}, we obtain a second   equation:
\begin{equation}\label{tauzz_1_alt}
\Big(\hat{\tau}_{zz} - \Upsilon_{sg}\mathcal{F}(\vec{\kappa}\cdot\hat{e}_z) - \frac{\Delta\Upsilon}{\pi}\frac{\sin s}{s}*\big(\mathcal{F}(\vec{\kappa}\cdot\hat{e}_z)\big)\Big)\Big|_{z=\tilde{h}} = N(s),
\end{equation}
where
\[ N(s) = \mathcal{F}(F_z) = \frac{2}{\sqrt{2\pi}}\Big(\frac{\gamma\sin\alpha}{R}\cos s-\Pi\frac{\sin s}{s}\Big). \]

While elasticity dominates the small wave number behavior of the displacements $u$ and $w$ in Fourier space, the traction stress generated by the geometry of the deformed substrate surface determines the decay of these transformed displacements for large wave numbers.  Including this traction stress is sufficient to influence the decay of these transformed displacements in Fourier space, thereby  eliminating the strain singularity at the contact line \cite{jerison11}.  The transformed terms in the curvature approximation  \eqref{curvature}   determine the dominant component   of the transformed displacements $\hat{u}$ and $\hat{w}$ as wave number $s \to \infty;$ these govern the decay estimates detailed in Appendix~\ref{sec:curvature}, which help justify an approximation of the traction stress:
\beq\label{curvature2}
\Upsilon(x)\vec{\kappa}(x) \approx k^2\frac{\Upsilon(x)}{R^2} \partial_{xx}u\hat{e}_x + \frac{\Upsilon(x)}{R^2}\partial_{xx}w\hat{e}_z.
\eeq

These terms are evaluated at the free surface $z=\tilde{h}$ and $k$ is the characteristic slope of the vertical displacement near the contact line.  Previous work \cite{bostwick14, jerison11, style12} has included the same vertical traction stress given by the second term in \eqref{curvature2}, but neglected the horizontal component (first term).
We find that the inclusion of the first term is sufficient to obtain bounded radial deformation at the contact line under the generalized contact line model, justified in \S3.3.  From the traction stress estimate \eqref{curvature2}, we calculate the transform estimates
\beq\label{curve_transforms}
\mathcal{F}(\vec{\kappa}\cdot\hat{e}_x) \approx -k^2\frac{s^2}{R^2}\hat{u}, \qquad \mathcal{F}(\vec{\kappa}\cdot\hat{e}_z) \approx -\frac{s^2}{R^2}\hat{w}.
\eeq

\subsection{Calculating Surface Deformation \label{sec:surfdef}}

Transforming the displacement and stress definitions in \eqref{dt}, equations \eqref{tauxz_1_alt} and \eqref{tauzz_1_alt} can now be written as a linear system for the unknowns $C(s), D(s):$ 
\begin{equation}\label{tauxz_2}
s^2C(s)\beta_{1}(s) + s^2D(s)\beta_{2}(s) + \frac{\Delta\Upsilon}{\pi}\frac{k^2}{R^2}\int_{-\infty}^\infty \frac{\sin(t-s)}{t-s} t^2\big(-i\hat{u}(t,\tilde{h})\big)dt = M(s),
\end{equation}
\begin{equation}\label{tauzz_2}
s^2C(s)\mu_{1}(s) + s^2D(s)\mu_{2}(s) + \frac{\Delta\Upsilon}{\pi}\frac{1}{R^2}\int_{-\infty}^\infty \frac{\sin(t-s)}{t-s}t^2 \big(\hat{w}(t,\tilde{h})\big)dt = N(s).
\end{equation}
The coefficient  functions $\beta_i$, $\mu_i$, $i=1,2$ are defined in Appendix~\ref{sec:transform}.  Solving  equations  \eqref{tauxz_2}, \eqref{tauzz_2} for   $C(s)$ and $D(s)$ then   defines the transform function $\hat{\psi}$ from which transform variables representing stresses $\hat{\tau}$ and displacements $\hat{u}$ are obtained.  

Since the displacement functions $u(x,z)$ and $w(x,z)$ given in \eqref{dt1} and \eqref{dt2} are second order derivatives of the potential function $\psi(x,z)$,  it follows that their transform variables $\hat{u}$ and $\hat{w}$ will be proportional to $s^2$.  Therefore, rather than approximating the Fourier coefficients $C$ and $D$ directly, we instead approximate $s^2 C$ and $s^2 D$.  Once the Fourier coefficients are calculated, the displacement field is obtained by approximating the inverse transforms derived from \eqref{dt1} and \eqref{dt2}:
\[ u(x,z) = \sqrt{\frac{2}{\pi}}\int_0^\infty \big(-i\hat{u}(s,z)\big)\sin(sx)ds \]
\begin{equation}\label{ft1}
 = \sqrt{\frac{2}{\pi}}\frac{1}{R^2}\int_0^\infty \Big[s^2C(s)\big(-sz\sinh(sz)\big)+s^2D(s)\big((3-4\nu)\sinh(sz)+sz\cosh(sz)\big)\Big]\sin(sx)ds\\
\end{equation}
\[ w(x,z) = \sqrt{\frac{2}{\pi}}\int_0^\infty \big(\hat{w}(s,z)\big)\cos(sx)ds \]
\begin{equation}\label{ft2}
 = \sqrt{\frac{2}{\pi}}\frac{1}{R^2}\int_0^\infty \Big[s^2C(s)\big(sz\cosh(sz)-(3-4\nu)\sinh(sz)\big)-s^2D(s)\big(sz\sinh(sz)\big)\Big]\cos(sx)ds
\end{equation}
These specify the displacement for $x \geq 0$, then odd and even extensions to $x<0$ are taken for $u$ and $w$ respectively.
We solve equations  \eqref{tauxz_2}, \eqref{tauzz_2} approximately by taking an asymptotic expansion of the transformed displacements $\hat{u}$, $\hat{w}$ as well as Fourier coefficients $C(s)$, $D(s)$:
\begin{subequations}\label{expand_CD}
\begin{eqnarray}
\hat{u}(s,z) &=& \hat{u}_0(s,z) + \epsilon \hat{u}_1(s,z) + \epsilon^2 \hat{u}_2(s,z) + \cdots,\\
\hat{w}(s,z) &=& \hat{w}_0(s,z) + \epsilon \hat{w}_1(s,z) + \epsilon^2 \hat{w}_2(s,z) + \cdots,\\
C(s) &=& C_0(s) + \epsilon C_1(s) + \epsilon^2 C_2(s) + \cdots,\\
D(s) &=& D_0(s) + \epsilon D_1(s) + \epsilon^2 D_2(s) + \cdots,
\end{eqnarray}
\end{subequations}
where we set the small parameter $\epsilon = \Delta\Upsilon/\Upsilon_{sg}$.  Under this expansion, we first solve for the zeroth order transformed displacements $\hat{u}_0$, $\hat{w}_0$ and use them to solve for the first order correction to the Fourier coefficients $C$, $D$.  The system of equations to obtain the first order correction is given by:
\begin{subequations}\label{IE_sys0}
\begin{eqnarray}
s^2C_0(s)\beta_1(s) + s^2D_0(s)\beta_2(s) &=& M(s),\\
s^2C_0(s)\mu_1(s) + s^2D_0(s)\mu_2(s) &=& N(s),
\end{eqnarray}
\end{subequations}
\begin{subequations}\label{IE_sys1}
\begin{eqnarray}
s^2C_1(s)\beta_1(s) + s^2D_1(s)\beta_2(s) &=& -\frac{\Upsilon_{sg}}{\pi}\frac{k^2}{R^2}\int_{-\infty}^\infty\frac{\sin(t-s)}{t-s}t^2\big(-i\hat{u}_0(t,\tilde{h})\big)dt,\\
s^2C_1(s)\mu_1(s) + s^2D_1(s)\mu_2(s) &=& -\frac{\Upsilon_{sg}}{\pi}\frac{1}{R^2}\int_{-\infty}^\infty\frac{\sin(t-s)}{t-s}t^2\big(\hat{w}_0(t,\tilde{h})\big)dt,
\end{eqnarray}
\end{subequations}

The zeroth order coefficients $s^2C_0$, $s^2D_0$ are algebraically obtained from \eqref{IE_sys0} and used to define the zeroth order transforms $\hat{u}_0$, $\hat{w}_0$.  The limits of the transformed displacements for large s values are given by
\begin{subequations}\label{trans_disp_lim}
\begin{eqnarray}
\hat{u}_0(s,\tilde{h}) &\sim& iEr_1\frac{\cos s}{s^3} + iEr_2\frac{\sin s}{s^2} = \hat{u}_{lim}(s),\\
\hat{w}_0(s,\tilde{h}) &\sim& Er_3\frac{\cos s}{s^2} + Er_4\frac{\sin s}{s^3} = \hat{w}_{lim}(s),
\end{eqnarray}
\end{subequations}
where constants $Er_j$, $j=1,2,3,4$ are given in Appendix B.  Substituting these limits into the right side of the equations in \eqref{IE_sys1} in place of the zeroth order transforms, we obtain the Cauchy principal values
\begin{equation}\label{Xres}
{\text{P.V.}}\hspace{.1in}\int_{-\infty}^\infty\frac{\sin(t-s)}{t-s}t^2\big(-i\hat{u}_{lim}(t)\big)dt = \frac{\pi}{2}Er_2\sin s,
\end{equation}
\begin{equation}\label{Zres}
{\text{P.V.}}\hspace{.1in}\int_{-\infty}^\infty\frac{\sin(t-s)}{t-s}t^2\big(\hat{w}_{lim}(t)\big)dt = \frac{\pi}{2}Er_3\cos s,
\end{equation}
which are obtained using contour integration in the complex plane.  To calculate the integrals in \eqref{IE_sys1}, the principal values \eqref{Xres} and \eqref{Zres} are added to the integral of the difference between the transformed displacements and their limits given in \eqref{trans_disp_lim}, calculated numerically over the interval $|s| \leq \bar{S}$.
 
This provides accurate results for the integrals in \eqref{IE_sys1} over the interval $|s| \leq S < \bar{S}$, where $S$ is the wave number cap for our calculations, the implications of which are examined in \S3.3.  With the solutions to \eqref{IE_sys1}, the first order transformed displacements $\hat{u}_1$, $\hat{w}_1$ are obtained and used to calculate the displacements $u(x,z)$, $w(x,z)$ by use of \eqref{ft1} and \eqref{ft2}.

We estimate the parameter $k$ using a method similar to a predictor-corrector procedure.  We first set $k = 0$ and calculate the derivative of the vertical displacement at the contact line   using the method above. This gives us a new value of $k$ from the formula
\beq\label{k_est}
k = \frac{1}{2R}\Big(\lim_{x \to 1^-}|\partial_xw(x,\tilde{h})| + \lim_{x \to 1^+}|\partial_xw(x,\tilde{h})|\Big).
\eeq

With this value of $k$, we recalculate the displacement field observing  that the vertical displacement is largely independent of $k$ near the contact line.  For the example scenario analyzed in \S\ref{sec:error} with parameters given in Fig.~\ref{error_plot}, $k$ was re-estimated with a relative error of $7.1 \times 10^{-7}$, illustrating the consistency and rapid convergence of the estimation \eqref{k_est}.

\subsection{Error Analysis \label{sec:error}}

\begin{figure}
\centerline{\includegraphics[scale=0.4]{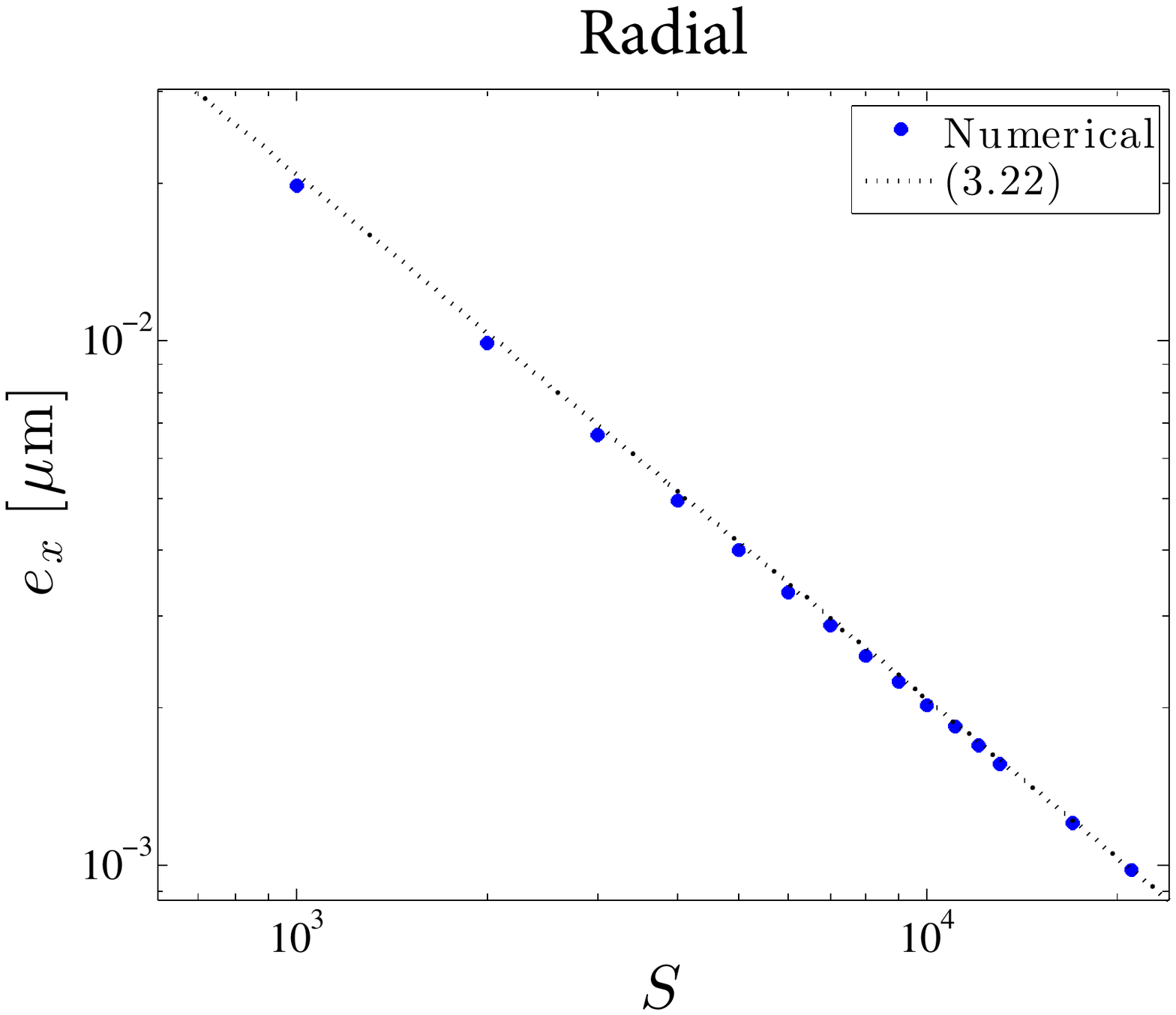}\hspace{.1in}
\includegraphics[scale=0.4]{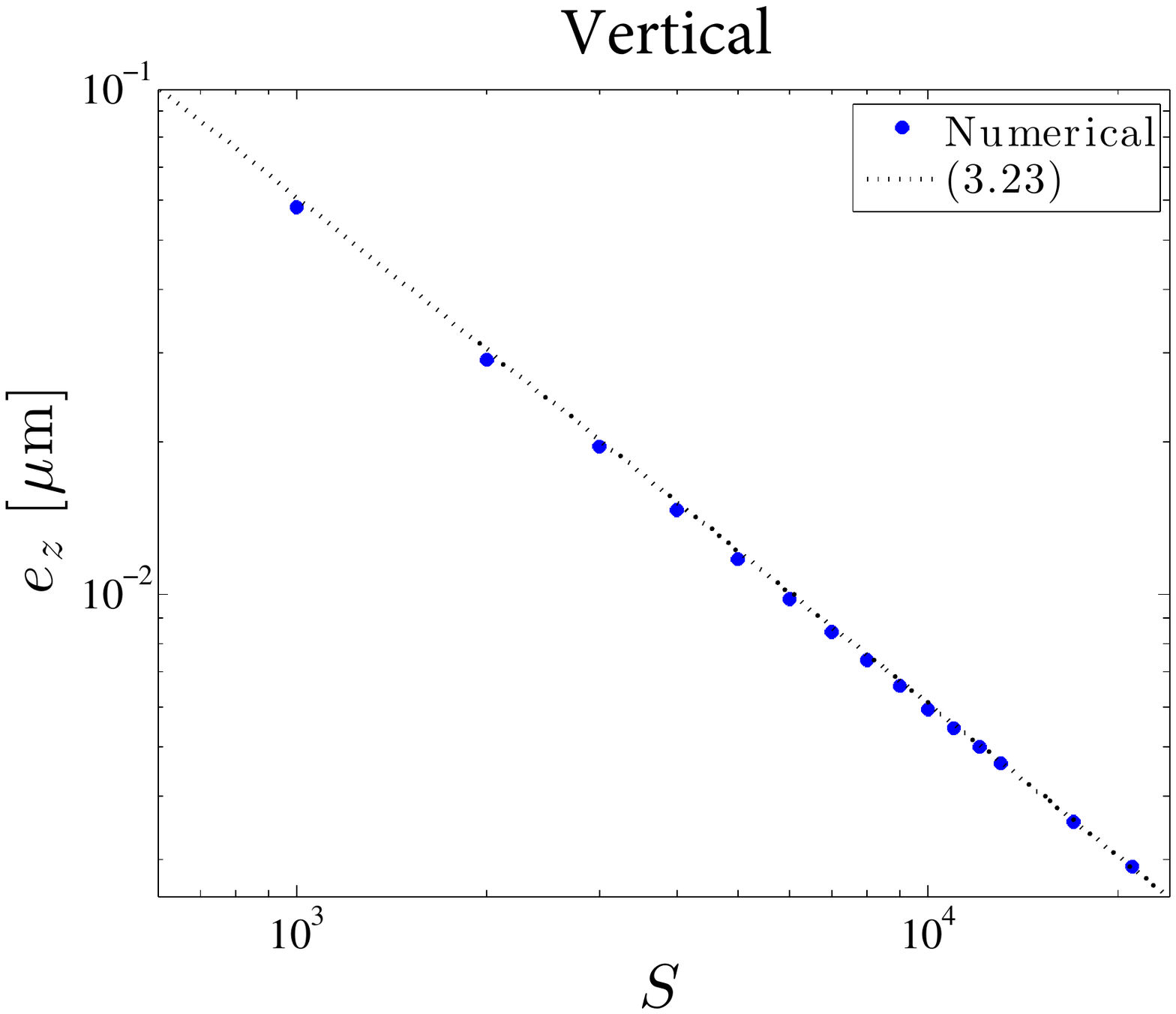}}
\caption{Radial (left) and vertical (right) numerical truncation error compared to predicted truncation error from \eqref{errx_2}, \eqref{errz_2} plotted vs wave number cap $S$ (log-log scale).  Parameters: $E=3$ kPa, $\nu=0.47$, $h=50$ $\mu$m, $R=150$ $\mu$m, $\gamma=50$ mN/m, $\Upsilon_{ls}=30$ mN/m, $\Upsilon_{sg}=42$ mN/m and $\bar{\Upsilon} = 36$ mN/m.  The numerical errors $e_x$ and $e_z$ decay according to \eqref{errx_2}, \eqref{errz_2}, confirming the deformation is bounded at the contact line.} 
\label{error_plot}
\end{figure}

In this section, we estimate the error in evaluating the inverse transforms only on the finite interval $0\leq s\leq S$.  In order to perform the estimate analytically, we consider only the simplified case in which $\bar{\Upsilon} = \Upsilon_{ls} = \Upsilon_{sg}$.  For comparison, we then simulate the numerical error for an example scenario where $\Upsilon_{ls} \neq \Upsilon_{sg}$. The two calculations are shown graphically  in Fig.~\ref{error_plot}.

We define the horizontal and vertical errors $e_x$, $e_z$ by
\beq\label{errx}
e_x = \max_x \Big|\sqrt{\frac{2}{\pi}}\int_S^\infty \big(-i\hat{u}(s,\tilde{h})\big)\sin(sx)ds\Big| 
\eeq
\beq\label{errz}
e_z = \max_x \Big|\sqrt{\frac{2}{\pi}}\int_S^\infty \big(\hat{w}(s,\tilde{h})\big)\cos(sx)ds\Big|
\eeq
With $s$ sufficiently large, we neglect the decay terms $e^{-s\tilde{h}}$ in the hyperbolic trig functions $\sinh(s\tilde{h})$ and $\cosh(s\tilde{h})$.  Moreover, we argue that the errors \eqref{errx}, \eqref{errz} are maximized at the contact line (details in Appendix~\ref{sec:curvature}).  As a result, we obtain the following   estimates as $S \to \infty$:
\beq\label{errx_2}
e_x \approx \Big|\sqrt{\frac{2}{\pi}}\int_S^\infty \frac{R^2M(s)}{k^2s^2\bar{\Upsilon}}\sin s \text{ }ds\Big| = \frac{2|F_r|R}{\pi k^2\bar{\Upsilon}}\int_S^\infty \frac{\sin^2s}{s^2} ds \approx \frac{|F_r|R}{\pi k^2\bar{\Upsilon}} \frac{1}{S},
\eeq
\beq\label{errz_2}
e_z \approx \Big|\sqrt{\frac{2}{\pi}}\int_S^\infty \frac{R^2N(s)}{s^2\bar{\Upsilon}}\cos s\text{ } ds\Big| \approx \frac{2\gamma\sin\alpha}{\pi \bar{\Upsilon}}\int_S^\infty \frac{\cos^2 s}{s^2} ds \approx \frac{\gamma R\sin\alpha}{\pi \bar{\Upsilon}} \frac{1}{S}.
\eeq

Vertical and radial displacement errors are calculated by recording the tip displacements relative to each other for different values of $S$ and are plotted in Fig.~\ref{error_plot} along with the error estimates from \eqref{errx_2} and \eqref{errz_2}.  The numerical simulations were calculated with a non-constant surface stress ($\Upsilon_{sg} \neq \Upsilon_{ls}$) using the mean solid stress as the characteristic stress $\bar{\Upsilon}$.  The error estimates reasonably fit the predicted error and we conclude that the deformation is bounded at the contact line by the inclusion of both traction boundary conditions.

\section{Results \label{sec:results}}
\begin{figure}
\centerline{\includegraphics[scale=0.4]{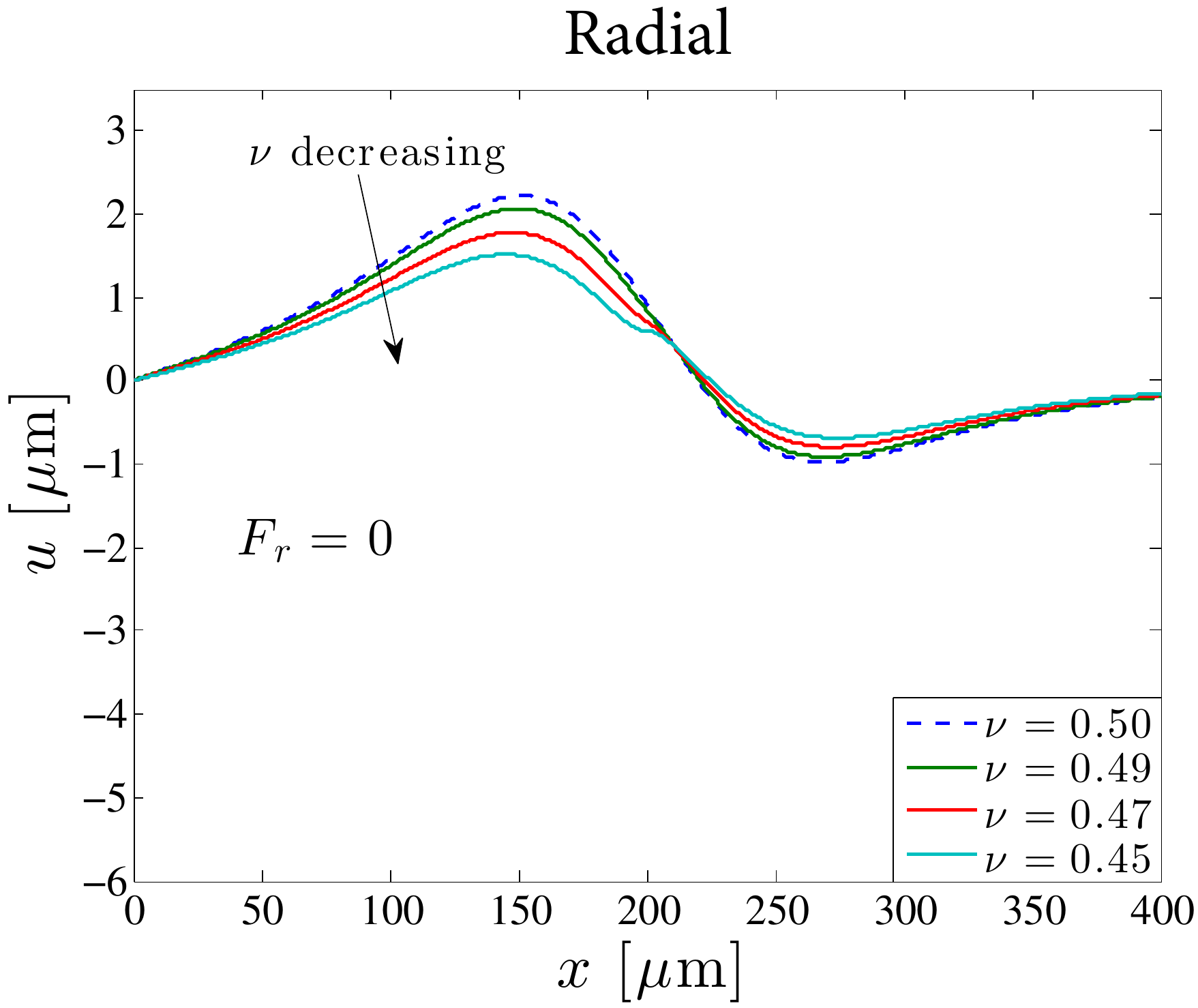}
\includegraphics[scale=0.4]{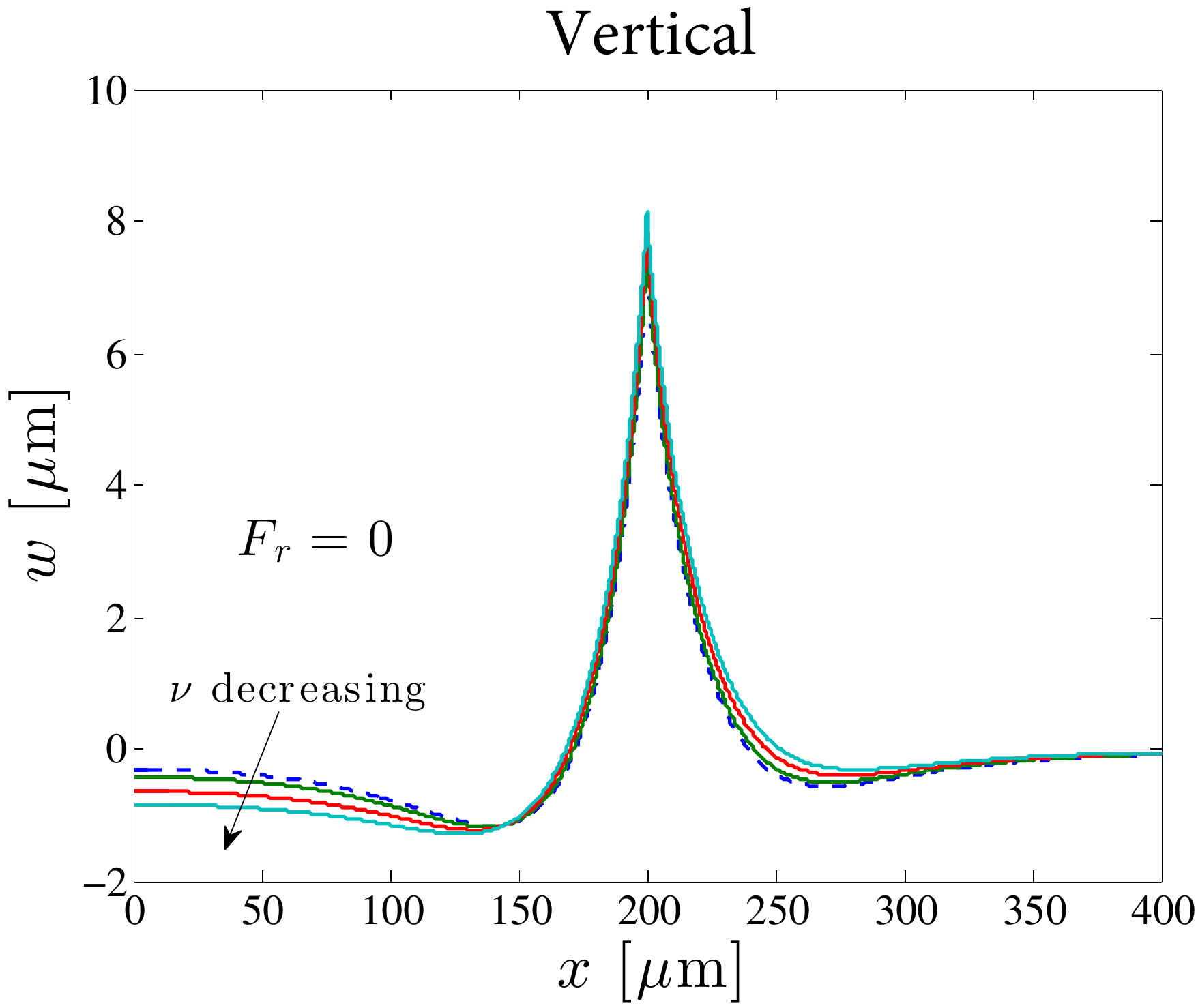}}
\vspace{-.15in}
\centerline{\includegraphics[scale=0.4]{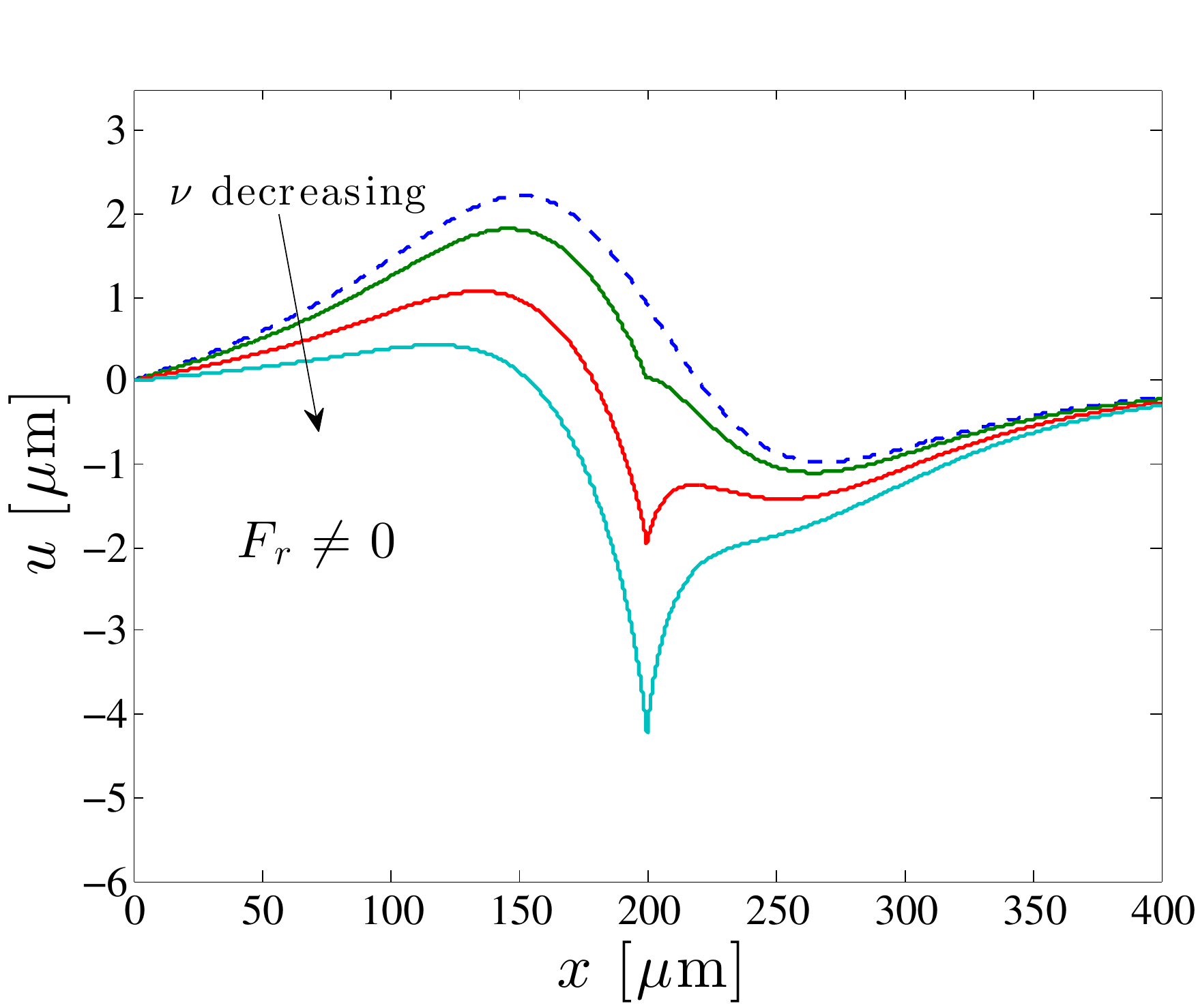}
\includegraphics[scale=0.4]{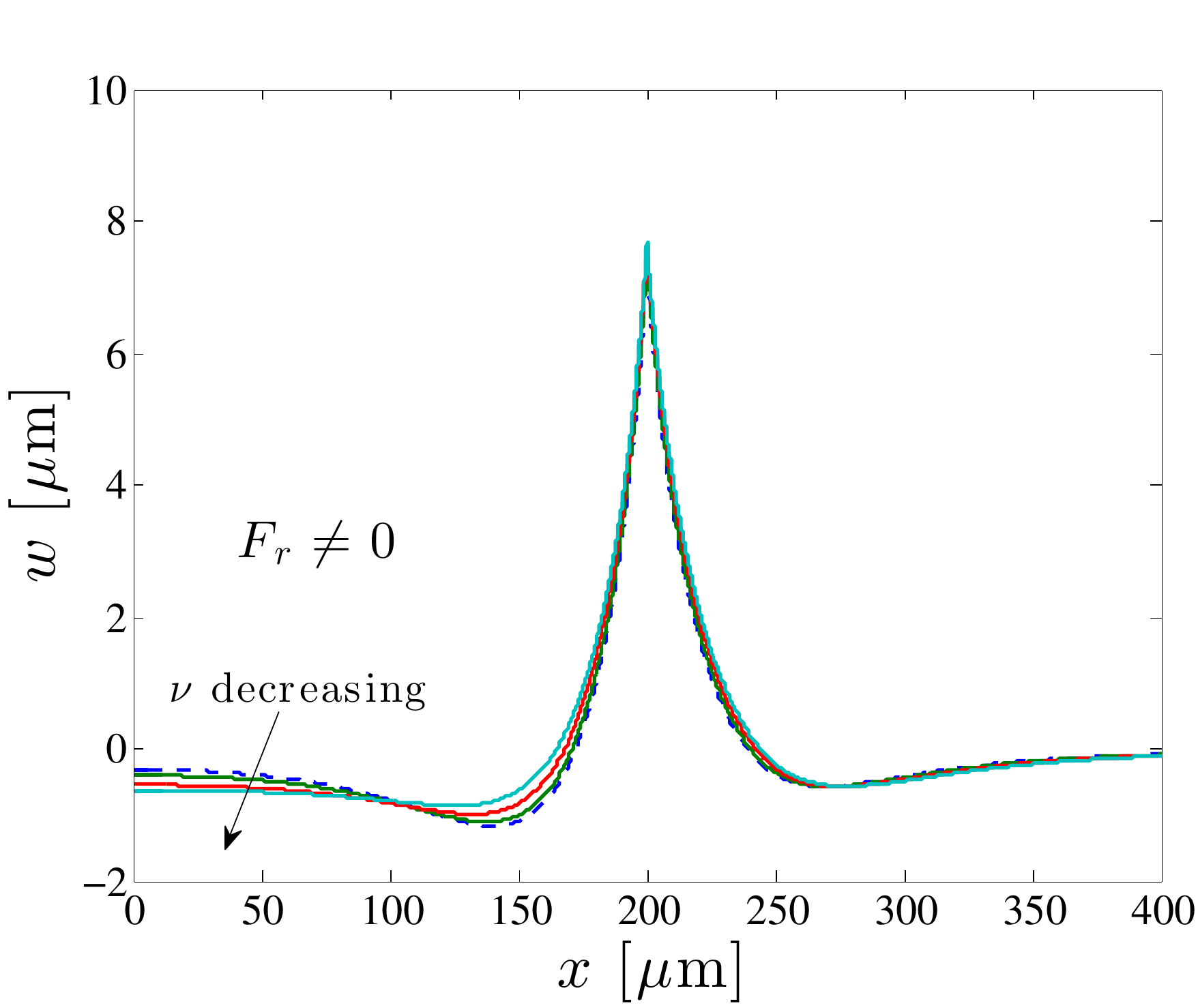}}
\caption{Radial (left) and vertical (right) displacements of the substrate surface for radial contact line force $F_r = 0$ mN/m (top) and $F_r = (46\text{ mN/m})\times(1-2\nu)/(1-\nu)(1+\cos\alpha)$ (bottom). Parameters: $E=4$ kPa, $h=50$ $\mu$m, $R=200$ $\mu$m, $\gamma=46$ mN/m, $\Upsilon_{ls}=33$ mN/m and $\Upsilon_{sg}=38$ mN/m.}
\label{Jfig}
\end{figure}

Using the method outlined in \S\ref{sec:solve}, we calculate both the radial ($u$) and vertical ($w$) surface displacements for an example substrate; this is shown in Fig.~\ref{Jfig} in physical units.  We include the displacements calculated with the conventional contact line model ($F_r = 0$) and the generalized contact line model with $F_r$ given by \eqref{J}.  
We vary the substrate Poisson's ratio $\nu$ to illustrate the effect of compressibility on the deformation of the substrate for both contact line models. Note that for the case of an incompressible substrate ($\nu = 1/2$), $F_r = 0$ regardless of the choice of contact line model, resulting in identical displacement calculations.  As seen in the radial displacements, the non-zero radial contact line force \eqref{J} pulls the substrate surface inward with increasing magnitude as the Poisson's ratio $\nu$ decreases.  This agrees with our intuition, as decreasing $\nu$ results in a larger radial contact line force predicted by \eqref{J}.

\begin{figure}
\centerline{\includegraphics[scale=0.4]{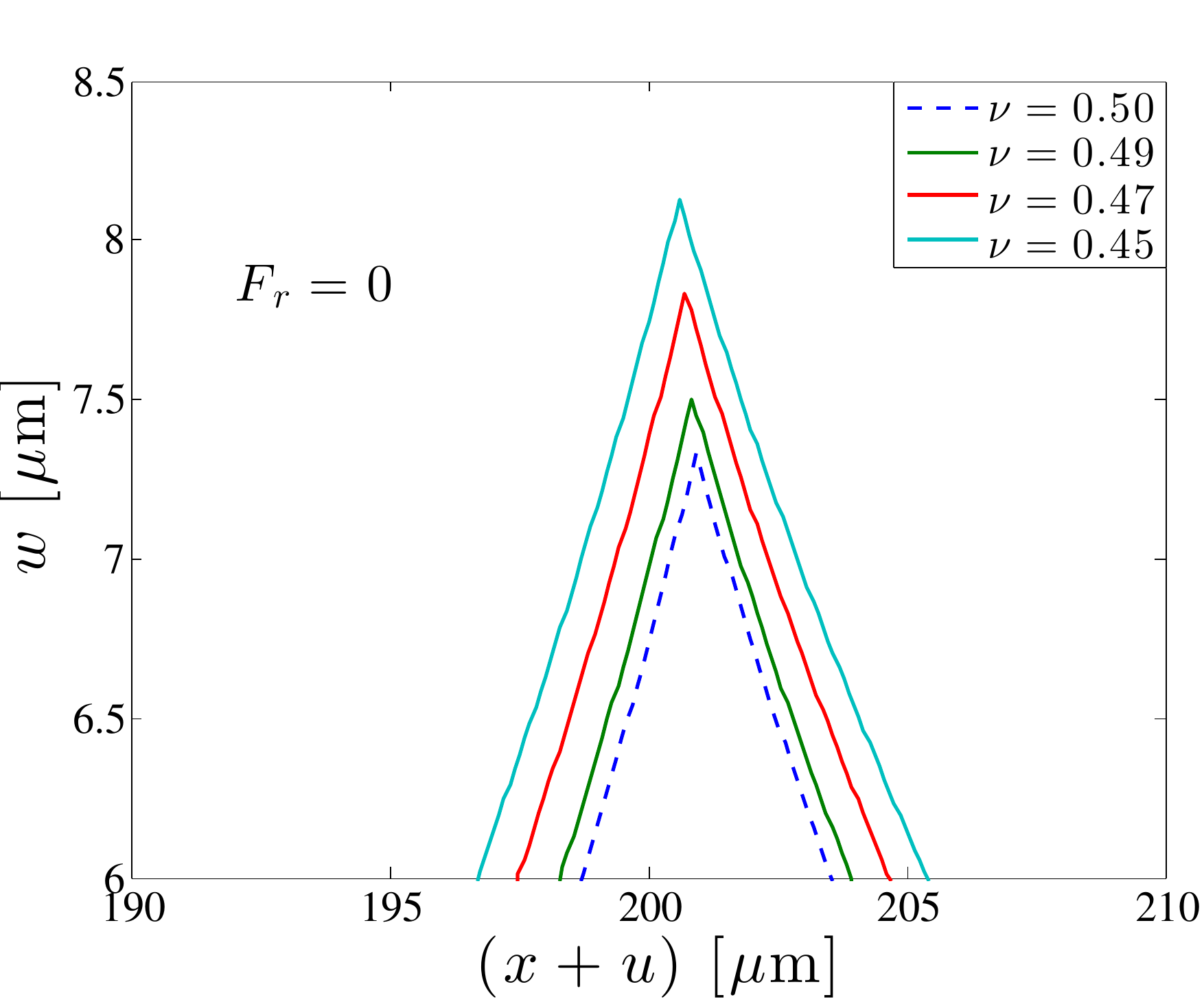}
\includegraphics[scale=0.4]{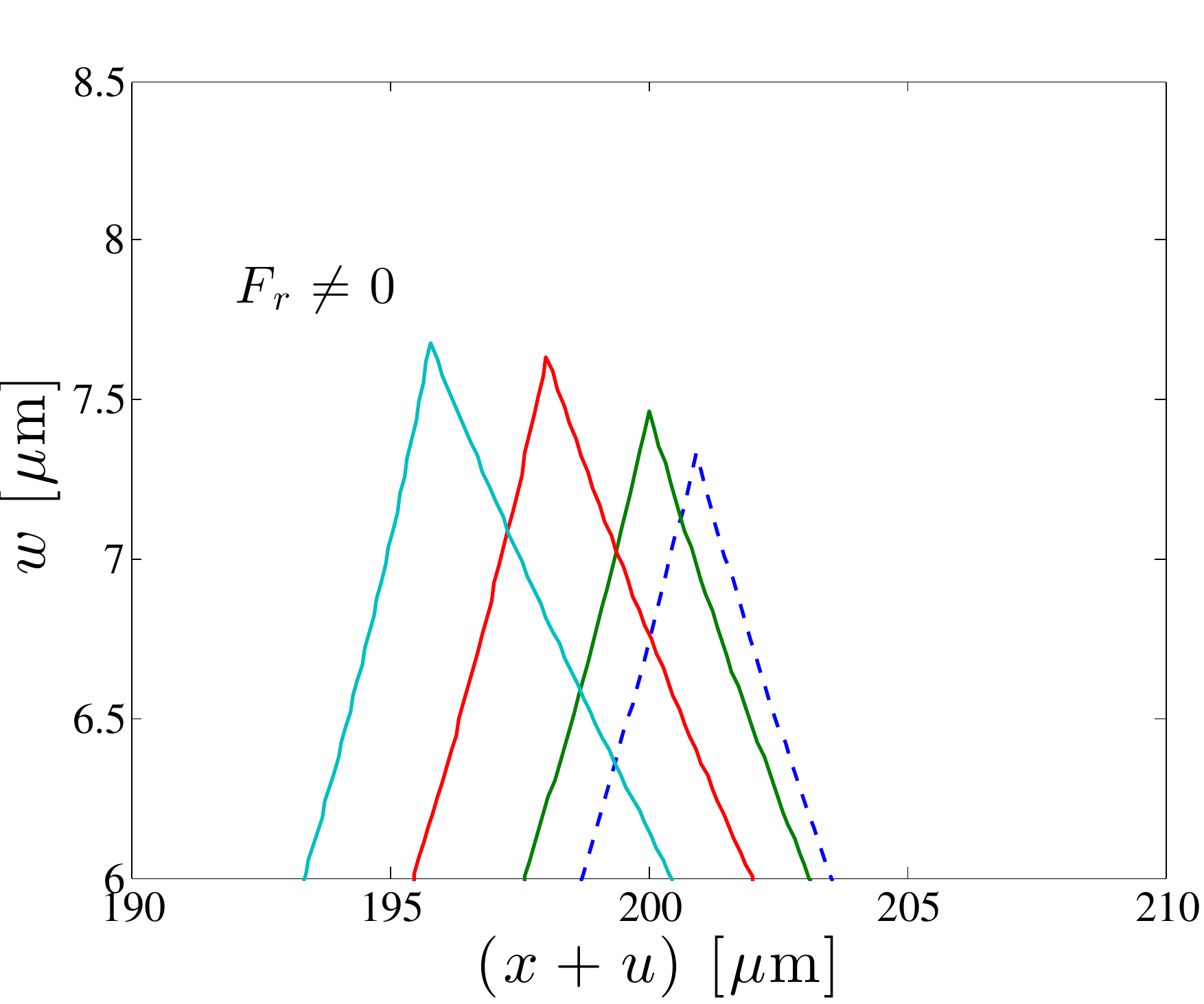}}
\caption{Surface deformation near the contact line for example substrate with radial contact line force $F_r = 0$ mN/m (left) and $F_r = (46\text{ mN/m})\times(1-2\nu)/(1-\nu)(1+\cos\alpha)$ (right).  Parameters identical to Fig.~\ref{Jfig}.}
\label{tip}
\end{figure}

In contrast to the radial displacement in Fig.~\ref{Jfig}, the vertical displacement results are very similar, regardless of contact line model.  Since the radial deformation is very small relative to the droplet radius $R$, the vertical displacement is a good visual representation of the actual substrate surface profile.  However, the radial deformation has a noticeable impact on the substrate surface profile near the contact line.  Combining the radial and vertical deformations gives us the parametrized substrate surface $\vec{r}(x) = \langle x+u,w+h\rangle_{z=h}$, which is shown in Fig.~\ref{tip} near the contact line for both models.  This illustrates the overall effect of the contact line force and the different behavior between the two contact line models.  Though not clearly visible in Fig.~\ref{Jfig}, the peak vertical displacement is lower for the generalized model ($F_r \neq 0$) for compressible substrates.  This is the result of a small decrease in the vertical contact line force for a non-trivial radial contact line force and constant solid surface stresses.  Using the definition of the contact angle given in \eqref{young1} and the radial force given in \eqref{J}, we obtain
\[ \cos\alpha = \frac{1-\nu}{\nu}\Big(\frac{\Upsilon_{sg} - \Upsilon_{ls}}{\gamma}\Big) + \frac{1-2\nu}{\nu}. \]
This equation increases in magnitude as $\nu$ is lowered in our simulations, resulting in a smaller vertical contact line force.

\section{Discussion \label{sec:discuss}}

We have presented a robust method for calculating the displacement field in a soft elastic substrate caused by a resting fluid droplet generalized to include partial wetting and (optionally) a radial contact line force.  In addition, from the framework presented, internal stresses can be analyzed within the substrate by constructing stress transform functions and approximating their respective inverse transforms.  The droplet shape has also been analyzed to provide a better understanding of how and when gravity affects the fluid pressure at the substrate surface.  

By including the previously-neglected radial traction stress, we gain a better understanding of the influence of the contact line in the radial direction.  Not incorporating this radial traction leads to a radial displacement whose transform decays as $\mathcal{O}(s^{-1})$ under a non-zero radial contact line force, which is insufficient to provide a bounded, realistic displacement.  The inclusion of this stress is critical in understanding the geometry of the contact line location, at which the force necessary to induce motion is transmitted to the droplet. 
Fig.~\ref{tip} provides quantitative predictions for what features to identify in future experiments.

With these advances in capability, we intend to use this model to further understand phenomena such as durotaxis, which depends largely on the displacement field and the contact angle of the droplet.  Durotaxis refers to the onset motion of a droplet caused by an underlying stiffness gradient in the substrate, and has been experimentally observed \cite{dufresne13} for droplets initially set on substrates with varying thickness.  The thickness of the substrate influences the local surface rigidity causing a change in contact angle across the droplet, which drives the motion of the droplet toward the softer (thicker) part of the substrate.  As simple as durotaxis is to describe, a purely mechanical model has yet to adequately replicate this phenomena.  Advancing our understanding of the deformation of the substrate surface for a substrate with varying rigidity will be critical in developing this model.

\appendix
\section{Gravitational Regime Pressure Limit \label{sec:gravity} }

To justify the limit in (\ref{pressure_lim}(b)), we first note that the integral \eqref{shape_sol} is singular when $\Pi/\rho gL_c=2\sin\frac \alpha 2,$ for which 
$\frac{f(0)}{L_c}=\Pi/\rho gL_c,$ its minimum value. Let $p=\Pi/\rho gL_c, $ and 
$p_0=2\sin \frac \alpha 2. $ Then $\frac{f(0)}{L_c}=p-\sqrt{p^2-p_o^2}$ is a function of $p$.  Moreover, the integral in \eqref{shape_sol} with $x=0$ that determines $R/L_c$ is also a function of $p,$ say $I(p)$. To find the asymptotic behavior of  $R/L_c$ as $p\searrow p_0,$ we find the leading order behavior of $I(p)$ as it approaches infinity.

Under the change of variables $\xi=p-\eta$, we obtain
\beq\label{integral1}
I(p)=\int^p_{q(p)} \frac{1-\frac 12(\eta^2-q^2) }{\sqrt{1-(\frac 12(\eta^2-q^2)-1)^2}}\, d\eta,     \quad q=q(p)=\sqrt{p^2-p_0^2}.
\eeq
Expanding the integrand $g(\zeta), \zeta=\sqrt{\eta^2-q^2},$ in powers of $\zeta,$ we find
\[ 
g(\zeta)\sim 1/\zeta +O(\zeta).
\]
Consequently, the leading order term in the integral \eqref{integral1} is
\[
\frac{R}{L_c}\sim\int_q^p\frac{d\eta}{\sqrt{\eta^2-q^2}}=\cosh^{-1}\frac pq.
\]
Solving algebraically for the pressure we obtain 
\beq\label{integral2}
\Pi\sim 2\rho gL_c\sin\frac \alpha 2\coth(R/L_c), \quad \mbox{as} \ R/L_c\to \infty. 
\eeq

\section{List of Transform Functions $\beta_j$, $\mu_j$ \& Constants $Er_j$ \label{sec:transform}}

\begin{subequations}\label{sys_coeff}
\begin{eqnarray}
\beta_1(s) &=& s\Big(\frac{(1-2\nu)E}{(1+\nu)R^3} - \frac{k^2s^2\tilde{h}\Upsilon_{sg}}{R^4}\Big)\sinh(s\tilde{h}) + s^2\Big(\frac{-E\tilde{h}}{(1+\nu)R^3}\Big)\cosh(s\tilde{h})\\
\beta_2(s) &=& s^2\Big(\frac{E\tilde{h}}{(1+\nu)R^3} + \frac{k^2(3-4\nu)\Upsilon_{sg}}{R^4}\Big)\sinh(s\tilde{h}) + s\Big(\frac{2(1-\nu)E}{(1+\nu)R^3} + \frac{k^2s^2\tilde{h}\Upsilon_{sg}}{R^4}\Big)\cosh(s\tilde{h})\\
\mu_1(s) &=& s^2\Big(\frac{E\tilde{h}}{(1+\nu)R^3} - \frac{(3-4\nu)\Upsilon_{sg}}{R^4}\Big)\sinh(s\tilde{h}) + s\Big(\frac{-2(1-\nu)E}{(1+\nu)R^3} + \frac{s^2\tilde{h}\Upsilon_{sg}}{R^4}\Big)\cosh(s\tilde{h})\\
\mu_2(s) &=& s\Big(\frac{-(1-2\nu)E}{(1+\nu)R^3} - \frac{s^2\tilde{h}\Upsilon_{sg}}{R^4}\Big)\sinh(s\tilde{h}) + s^2\Big(\frac{-E\tilde{h}}{(1+\nu)R^3}\Big)\cosh(s\tilde{h})\\
Er_1 &=& \frac{2(1-2\nu)R^2E\gamma\sin\alpha}{\sqrt{2\pi}(3-4\nu)(1+\nu)k^2\Upsilon_{sg}^2}\\
Er_2 &=& \frac{-2RF_r}{\sqrt{2\pi}k^2\Upsilon_{sg}}\\
Er_3 &=& \frac{2R\gamma\sin\alpha}{\sqrt{2\pi}\Upsilon_{sg}}\\
Er_4 &=& \frac{-2R^2}{\sqrt{2\pi}(3-4\nu)k^2\Upsilon_{sg}^2}\Big((3-4\nu)k^2\Pi\Upsilon_{sg} + \frac{(1-2\nu)EF_r}{(1+\nu)}\Big)
\end{eqnarray}
\end{subequations}

\section{Curvature Approximation and Detailed Error Analysis \label{sec:curvature}}

For the curvature and error analysis, we assume that the solid surface stress is constant ($\Upsilon(x)  = \bar{\Upsilon}$).
This reduces the equations \eqref{tauxz_1_alt}, \eqref{tauzz_1_alt} to 
\begin{equation}\label{sys1_2}
-i\Big(\hat{\tau}_{xz} - \bar{\Upsilon}\mathcal{F}\big(\vec{\kappa}\cdot\hat{e}_x\big)\Big)\Big|_{z=\tilde{h}} = M(s)\end{equation}
\begin{equation}\label{sys2_2}
\Big(\hat{\tau}_{zz} - \bar{\Upsilon}\mathcal{F}\big(\vec{\kappa}\cdot\hat{e}_z\big)\Big)\Big|_{z=\tilde{h}} = N(s)
\end{equation}
Agreeing with \cite{bostwick14} regarding the vertical component of curvature, we take
\[ \mathcal{F}(\vec{\kappa}\cdot\hat{e}_z) \approx \frac{1}{R^2}\mathcal{F}(\partial_{xx}w) = -\frac{s^2}{R^2}\hat{w} \]
while the FT of the radial curvature component is then taken to be
\begin{equation}\label{kappa_x}
\mathcal{F}\big(\vec{\kappa}\cdot\hat{e}_x\big) = \frac{1}{R^2}\mathcal{F}\Big(\frac{(\partial_xw)^2}{R^2}\partial_{xx}u\Big) - i\frac{s}{R^3}\Big[s^2\mathcal{F}\Big(\frac{w^2}{4}\Big) + \mathcal{F}\Big(\frac{w\partial_{xx}w}{2}\Big)\Big].
\end{equation}
Approximations are taken in equation \eqref{kappa_x} by acknowledging that the second derivatives of the displacement are large near the contact line where the sharp kink is located, but by comparison are negligibly small elsewhere in the spatial domain.  Therefore the largest contribution to the transforms involving second derivatives comes from near the contact line location.  For the transform of $\mathcal{F}(w^2/4)$, we take a constant value approximation for $w$ near the contact line and leave the rest in the transform.  For the transforms with second derivatives of displacements $u$ and $w$, the remaining terms are approximated by a constant value.  We take $k^2 \approx \big(\partial_xw(1,\tilde{h})\big)^2/R^2$ and $U \approx w(1,\tilde{h})$ and the transform \eqref{kappa_x} becomes
\[ \mathcal{F}(\vec{\kappa}\cdot\hat{e}_x) \approx -k^2\frac{s^2}{R^2}\hat{u} + i\frac{U}{4}\frac{s^3}{R^3}\hat{w}. \]
With the curvature transform approximations, the system of equations given by \eqref{sys1_2} and \eqref{sys2_2} becomes
\begin{equation}\label{sys1_3}
s^2C(s)\beta^*_1(s) + s^2D(s)\beta^*_2(s) = M(s)
\end{equation}
\begin{equation}\label{sys2_3}
s^2C(s)\mu_1(s) + s^2D(s)\mu_2(s) = N(s)
\end{equation}
where
\[ \beta^*_1(s) = \Big(\frac{(1-2\nu)Es}{(1+\nu)R^3} + U\frac{(3-4\nu)\bar{\Upsilon} s^3}{4R^5} - k^2\frac{\bar{\Upsilon}\tilde{h}s^3}{R^4}\Big)\sinh(s\tilde{h}) + \Big(-\frac{E\tilde{h}s^2}{(1+\nu)R^3} - U\frac{\bar{\Upsilon}\tilde{h}s^4}{4R^5}\Big)\cosh(s\tilde{h}) \]
\[ \beta^*_2(s) = \Big(\frac{E\tilde{h}s^2}{(1+\nu)R^3} + U\frac{\bar{\Upsilon}\tilde{h}s^4}{4R^5} + k^2\frac{(3-4\nu)\bar{\Upsilon} s^2}{R^4}\Big)\sinh(s\tilde{h}) + \Big(\frac{2(1-\nu)Es}{(1+\nu)R^3} + k^2\frac{\bar{\Upsilon}\tilde{h}s^3}{R^4}\Big)\cosh(s\tilde{h})\] 
and $\mu_1$, $\mu_2$ are given in Appendix~\ref{sec:transform}.
Solving the system \eqref{sys1_3}, \eqref{sys2_3} for Fourier coefficients $s^2C(s)$ and $s^2D(s)$ algebraically yields
\begin{equation}\label{CD}
s^2C(s) = \frac{\mu_2(s)M(s) - \beta^*_2(s)N(s)}{\chi(s)} \qquad s^2D(s) = \frac{\beta^*_1(s)N(s) - \mu_1(s)M(s)}{\chi(s)}
\end{equation}
where
\[ \chi(s) = \beta^*_1(s)\mu_2(s) - \beta^*_2(s)\mu_1(s).\]
Elastic terms introduced by the transforrmed stresses $\hat{\tau}$ dominate the traction terms introduced by the curvature in the transformed displacements $\hat{u}$ and $\hat{w}$ for low wave numbers ($s/R \ll 1 \mu$m$^{-1}$).  We assign a wavenumber $S_T = \mathcal{O}(R)$ where the traction terms begin to dominate the elastic terms in the displacement calculations.  We define the displacement from wave numbers in this regime the traction displacements:
\[ u^T(x) = \sqrt{\frac{2}{\pi}}\int_{S_T}^\infty \hat{u}(s,\tilde{h})\big(-i\sin(sx)\big)ds \]
\[ w^T(x) = \sqrt{\frac{2}{\pi}}\int_{S_T}^\infty \hat{w}(s,\tilde{h})\big(\cos(sx)\big)ds \]
With the definitions of displacement variables in $\eqref{ft1}$ and $\eqref{ft2}$, and the algebraic representation of the Fourier coefficients from \eqref{CD}, we can obtain an algebraic representation of the integrands.  We then acknowledge that the hyperbolic trig functions are asymptotically similar in the range of wave numbers used to calculate the traction displacements.  We are then left with rational functions multiplied by the oscillatory contact line influence terms $M$ and $N$ and the oscillatory transform terms $\sin$ and $\cos$.  We acknowledge that these trig functions are orthogonal, and therefore the largest contribution of these traction displacements will take place near the contact line where the oscillatory terms dominantly interact.  We assume they are maximized at the contact line itself ($x=1$), which gives us
\begin{equation}\label{traction3}
\max_x |u^T| \approx \frac{2|F_r|}{\pi R^3}\Big|\int_{S_T}^\infty \frac{2(1-\nu)(3-4\nu)A_1(s) + (3-4\nu)^2A_3(s)}{\tilde{\chi}(s)} \sin^2(s) ds\Big|  = \bar{u}^T
\end{equation}
\begin{equation}\label{traction4}
\max_x |w^T| \approx \frac{2\gamma\sin\alpha}{\pi R^3}\Big|\int_{S_T}^\infty \frac{2(1-\nu)(3-4\nu)A_1(s) + k^2(3-4\nu)^2A_3(s)}{\tilde{\chi}(s)} \cos^2(s) ds\Big| = \bar{w}^T
\end{equation}
where
\[ A_1(s) = \frac{Es}{(1+\nu)R^3},\qquad A_2(s) = \frac{\bar{\Upsilon} s^3}{R^5}, \qquad A_3(s) = \frac{\bar{\Upsilon} s^2}{R^4}\]
and
\[ \tilde{\chi}(s) = (3-4\nu)A_1^2(s) - \frac{U}{4}(1-2\nu)(3-4\nu)A_1(s)A_2(s) + 2(k^2+1)(1-\nu)(3-4\nu)A_1(s)A_3(s) + k^2(3-4\nu)^2A_3^2(s) \] 
Here we note that the influence of parameter $U$ is completely eliminated from the numerator.  We then analyze the highest order terms of the denominator $\tilde{\chi}$ as
\[ \tilde{\chi}(s) \sim\frac{(3-4\nu)\bar{\Upsilon}^2}{R^8}\Big((3-4\nu)k^2 - \frac{1-2\nu}{4(1+\nu)}\frac{U}{\bar{\Upsilon}/E}\Big)s^4 \]
Assuming $U$ is on the order of the elastocapillary length scale $L_e = \gamma/E$, the ratio of the two terms becomes
\[ \Big(4(1+\nu)k^2\frac{\bar{\Upsilon}/E}{U}\Big)\frac{3-4\nu}{1-2\nu} = \mathcal{O}\Big(\frac{3-4\nu}{1-2\nu}\Big) \gg 1 \]
for nearly incompressible substrates.  Approximating the integrals by taking the leading order term of the numerator and denominator, we have that the parameter $U$ can be removed from the system, giving the $\beta$ functions used in the paper (shown in Appendix~\ref{sec:transform}).  This gives us the simplified traction stress estimate \eqref{curvature2}.  Approximating the integrals \eqref{traction3} and \eqref{traction4} by using the mean value of the squared trig functions gives the error estimates presented in \S\ref{sec:error} after substituting our wave number cap $S$ for the traction wave number $S_T < S$.

We also note that removing the radial traction from these estimates, or equivalently setting $k^2 \equiv U \equiv 0$, provides an unbounded radial traction displacement $\bar{u}^T$.  Let us call $\bar{\bar{u}}^T$ and $\bar{\bar{w}}^T$ the traction displacement estimates with no radial traction boundary condition.  We find that $\bar{\bar{w}}^T \sim \bar{w}^T$ but the integrand needed to calculate $\bar{\bar{u}}^T$, decays at a rate of $\mathcal{O}(s^{-1})$, which makes the radial traction displacement estimate to be unbounded without the radial traction boundary condition.  Further detail for material in Appendix~\ref{sec:curvature} can be found in the supplementary material.

\acknowledgements
This work was supported by the National Science Foundation under grant number DMS-1517291.  We would like to thank Josh Bostwick for discussions regarding the solution method.  We would also like to thank participants of the Capillarity of Soft Interfaces conference of November 2015 hosted at the Lorentz Center in Leiden, NL with whom discussions were held; namely Bruno Andreotti, Jacco Snoeijer, Eric Dufresne, Robert Style, Stefan Karpitschka, Laurent Limat and Anand Jagota.

\label{lastpage}
\end{document}